\documentclass[twocolumn,showpacs,preprintnumbers,amsmath,amssymb,10pt,prd]{revtex4}

\usepackage{graphicx}
\usepackage{dcolumn}
\usepackage{bm}
\usepackage{hyperref}
\usepackage[pdftex]{color}
\usepackage{mathrsfs}

\def\Hc{\mathscr{H}}
\def\Pl{{_{\!P\!\ell}}}
\newcommand{\lb}[1]{\label{#1}}

\newcommand{\bark}{\bar{k}}
\newcommand{\barp}{\bar{p}}
\newcommand{\barmu}{\bar{\mu}}
\newcommand{\barn}{\bar{N}}
\newcommand{\barN}{\bar{N}}

\newcommand{\barpi}{\bar{\pi}}

\newcommand\be{\begin{equation}}
\newcommand\ee{\end{equation}}
\newcommand\ba{\begin{eqnarray}}
\newcommand\ea{\end{eqnarray}}

\begin{document}
\title{Consistency of holonomy-corrected scalar, vector and tensor perturbations
\\ in Loop Quantum Cosmology}

\author{Thomas  Cailleteau, Aur\'elien Barrau and Francesca Vidotto}%
\affiliation{%
Laboratoire de Physique Subatomique et de Cosmologie, UJF, INPG, CNRS, IN2P3\\
53, avenue des Martyrs, 38026 Grenoble cedex, France
}

\author{Julien Grain}
 \affiliation{%
Institut\,d'Astrophysique\,Spatiale, Universit\'e\,Paris-Sud, CNRS\\
Batiments 120-121, 91405 Orsay cedex, France
}

\date{\today}

\begin{abstract}

Loop Quantum Cosmology yields two kinds of quantum corrections to the effective equations of motion for cosmological perturbations. Here we focus on the holonomy kind and we study the problem of the closure of the resulting algebra of constraints. 
Up to now, tensor, vector and scalar perturbations were studied independently, leading to different algebras of constraints. The structures of the related algebras were imposed by the requirement of anomaly freedom. In this article we show that the algebra can be modified by a very simple quantum correction, holding for all types of perturbations.  This demonstrates the consistency of the theory and shows that lessons from the study of scalar perturbations should be taken into account when studying tensor modes.  The Mukhanov-Sasaki equations of motion are similarly modified by a simple term.

\end{abstract}


\maketitle

\paragraph*{\bf Introduction--}
Loop Quantum Gravity (LQG) is a promising framework for a background-invariant 
non-perturbative quantization of general relativity (GR)
-- see \cite{lqg_review} for introductory reviews. 
The theory can be derived from different paths, going from a formal quantization of 
geometry to a covariant or canonical quantizations of GR, all yielding to the same 
theory. 
In the canonical formulation, the loop quantization is
obtained by choosing the holonomy of the gravitational connection and the 
flux of the densitized triad as basic variables. 
Loop quantum cosmology (LQC) is the symmetry reduced version of LQG. 
Although a rigorous complete derivation from the full theory is still missing,
LQC utilizes key elements of LQG for studying quantum corrections
of the cosmological dynamics. These corrections turn out to be negligible at 
low curvature, and important where the energy density approaches the 
Planck scale $\rho_\Pl$. They give rise to a  strong effective repulsive force which
replaces the big bang by a big bounce (see {\it e.g.} \cite{lqc_review} for a review). 

As for any tentative theory of quantum gravity, experimental tests are still missing, and
searching for observational signatures is obviously a key
challenge. Cosmological perturbations, which are directly related to measurable 
spectra, provide the best link to observation. 
Here we consider the theory of \emph{linear} cosmological perturbations in the Hamiltonian framework \cite{Langlois:1994ec}.
The theoretical analysis of these 
perturbations can be guided by a consistency requirement: the absence
of anomalies that would jeopardize the closure of the effective constraint algebra.
This requirement has been so far separately analyzed for scalar, vector, 
and tensor perturbations, leading to different corrections to the constraints.
This work focuses on the issue of finding a unique self-consistent algebra of constraints 
making the approach consistent for any kind of perturbation. We present a consistent constraint 
structure suitable for all types of perturbations, and leading to a simple modification of the gauge-invariant Mukhanov-Sasaki equation of motion.  This shows  the overall consistency of the theory 
and indicates that results of the analysis of the scalar perturbations must be taken into account to study
tensor modes.\\

\paragraph*{\bf Theoretical framework--}
LQC is formulated in the canonical language. Because of general covariance the
canonical Hamiltonian is a combination of constraints  $\mathcal{C}_I$. Consistency requires 
that the constraints are preserved under the evolution
they generate. This is assured in the classical theory by the closure of the Poisson algebra of the 
constraints 
\begin{equation}
\{ \mathcal{C}_I, \mathcal{C}_J \} = {f^K}_{IJ}(A^j_b,E^a_i) \mathcal{C}_K, \label{algebra}
\end{equation}
where $\mathcal{C}_I$, $I=1,2,3,$ are the Gauss, diffeomorphism and Hamiltonian constraints and
${f^K}_{\!IJ}( A^j_b,E^a_i)$ are structure functions which, in general, 
depend on the phase space (Ashtekar) variables  $(A^j_b, E^a_i)$.
In LQC, quantum corrections can be studied as effective modifications of the 
Hamiltonian constraint. In doing so, anomalies generically appear: the modified 
constraints $\mathcal{C}^Q_I$ do not
form  a closed algebra anymore:
\begin{equation}
\{ \mathcal{C}^Q_I, \mathcal{C}^Q_J \} = {f^K}_{IJ}(A^j_b,E^a_i)\, \mathcal{C}^Q_K+
\mathcal{A}_{IJ}. 
\end{equation}
The anomalous term $\mathcal{A}_{IJ}$ can be removed by  carefully adjusting the form of the quantum correction to the Hamiltonian constraint. This is achieved by adding suitable ``counterterms" that vanish in the classical limit. The resulting deformed algebra can be phenomenologically very rich. 

In the case of a \emph{flat} FLRW background, the Ashtekar variables
can be decomposed as follows
\begin{equation}
 A^i_a = \gamma \bar{k} \delta^i_a +\delta A^i_a \ \ \ {\rm and} \ \ \ E_i^a = \bar{p} \delta_i^a +\delta E_i^a, 
 \label{tre}
\end{equation}
where $\bar{k}$ and $\bar{p}$ parametrize the background phase space, and 
$\gamma$ is the Barbero-Immirzi parameter. 
The variation of the connection receives contributions from the fluctuations of both the intrinsic and extrinsic curvature: $\delta A^i_a=\delta \Gamma^i_a+\gamma\delta K^i_a$.

LQC generates two main classes of effective corrections to the constraints, called the inverse-volume corrections and the holonomy corrections \cite{lqc_review}. 
The closure of the algebra of cosmological perturbations has been extensively 
studied for  inverse-volume corrections. It was explicitly shown that closure 
can indeed be achieved. This was demonstrated for scalar \cite{Bojowald:2008gz,Bojowald:2008jv}, 
vector \cite{Bojowald:2007hv} and tensor modes \cite{Bojowald:2007cd}. 
Using the anomaly-free scalar perturbations, predictions 
for the power spectrum  were also obtained  
\cite{Bojowald:2010me}. This allowed to put constraints on 
some parameters of the model using observations of the cosmic microwave 
background radiation (CMB) \cite{Bojowald:2011hd}.  

Here, we focus on the holonomy corrections -- appearing because of the use of the holonomy of the Ashtekar
connection. It is worth
emphasizing that for tensor modes, the algebra is automatically anomaly-free. For this reason, several
works were devoted to the phenomenology of holonomy-corrected tensor perturbations (see {\it e.g.}
\cite{pheno_tensor}). The anomaly-free algebra for vector modes was studied in \cite{Li:2011zzd} and
recently fully derived, including matter, in \cite{vector}. The scalar algebra was obtained 
in \cite{scalar}.\\

\paragraph*{\bf Perturbations--}
Taking into account the form of the perturbed variables \eqref{tre},
we introduce a general expression for the variation of the spin connection as 
\ba \lb{deltaGamma}
\delta \Gamma^i_a &=&  \frac{1}{2 \barp}\   X^{ijb}_{ca}\   \partial_b \delta E^c _j + \frac{1}{2 \barp^2} \ Y^{ijkl}_{abc}\  \delta E^b_j \partial_k \delta E^c_l, \\
 \lb{Xijbca}\mbox{where}&&
X^{ijb}_{ca} =  \epsilon^{ij}_c \delta ^b_a - \epsilon^{ib}_c \delta ^j_a + \epsilon^{ijb} \delta_{ca} + \epsilon^{ib}_a \delta ^j_c.
\hskip15mm
\ea
$Y^{ijkl}_{abc}$ has an expression similar to $X^{ijb}_{ca}$, but more complicated: 
it is not needed here explicitly, because it appears
only as a boundary term in the second order term of the Hamiltonian constraint \eqref{annulation} in a way that does
not affect the equations of motion. 
The information about what kind of perturbations we consider
(scalar, vector or tensor perturbations) is coded in the term $ \frac{1}{2 \barp}\, X^{ijb}_{ca}\, \partial_b \delta E^c _j$. 

The variation of the densitized triad can be decomposed as follows:
\ba
\delta E^a_i &=& \barp \;\Big[ - 2 \psi \delta^a_i + (\delta^a_i \partial^d \partial_d  - \partial^a \partial_i )E 
\nonumber\\ && 
- c_1 \partial^a F_i - c_2 \partial_i F^a - {\scriptstyle\frac{1}{2}} h^a_i \; \Big],
\ea
where the first two terms $\psi$  and  $E$ correspond to scalar modes, the terms with $F_i$ and $F^a$ to vector modes and the term with $h_i^a$ to the tensor mode.
Vector modes are transverse, and tensor modes are transverse and traceless. These conditions constrain $\delta E^a_i$ and  $\delta K^i_a$, as well as the lapse $\delta N$ and the shift vector $\delta N^a$.  In particular, vanishing trace implies
\be
\delta^i_a \delta E^a_i =\delta ^a_i \delta K^i_a= 0~.
\ee
Tensor and vector perturbations satisfy this condition, so that in these cases the terms
containing these expressions disappear from the constraints. The form of the metric in the case of vector and tensor modes implies that the variation of the lapse is zero: $\delta N=0$\,. Therefore, some first-order constraints do not influence the perturbed dynamics.

For vector modes, the variation of the shift corresponds to one of the two degrees of freedom indicated with $S_a$ and $F_a$: $\delta N^a = S^a$.
For tensor modes instead, the transverseness, {\it i.e.} null divergence, implies
\be
\partial^ i\delta E^a_i =
\partial_a \delta E^a_i = 0~.
\ee
As above, the form of the metric for tensor modes implies $\delta N^a=0$ for the shift, so that some further first-order terms do not contribute to the dynamics.

Scalar perturbations are the more general: no term disappears and all the constraints contribute to define the perturbed dynamics. We have
\be \lb{pertscaldeltaN} \!
 \delta N = \barN \phi \hskip 5mm \mbox{and}  \hskip 5mm \delta N^a = \partial^a B, \!\!
\ee
where $\barN$ is the unperturbed part of the lapse $N=\barN+\delta N$ and $\phi$ and $B$ are scalar fields. 
\\[1em]
If  we turn-on the quantum corrections by modifying the Hamiltonian constraint, anomalies appear and we have to add counterterms in order to make the Poisson algebra closed. In previous works, these counterterms were found considering separately the case of each kind of modes. The tensor and vector cases were simpler because of the vanishing of several terms in the constraints, as observed. The scalar case, on the other hand, is from this perspective the most general one, since all the constraint terms are present.  It is indeed easy to see that the counterterms that adjust the Hamiltonian for the scalar case \cite{scalar} work \emph{also} for the vector and tensor cases, thus providing a general solution to the closure of the algebra. 
Therefore starting from the scalar case it is possible to define a unique closed algebra of modified constraints, with the most general counterterms, giving back  correct counterterms for scalar and tensor perturbations when imposing transverseness and vanishing trace.
\\

\paragraph*{\bf Constraints--}
We consider the algebra of the diffeomorphism and Hamiltonian constraints (see \cite{lqc_review} for the expression of the constraints in terms of the variables \eqref{tre}).  In each constraint, gravity and matter -- here modeled by a single scalar field with canonical variables $(\varphi,\pi)$ -- contribute. 
\paragraph*{Diffeomorphism constraint--}
The diffeomorphism constraint can be decomposed as
\be 
D[N^a] = \int_\Sigma d^3x \left[ \barn^a\,( \mathcal{D}^{(0)}+ \mathcal{D}^{(2)}) +  \delta N^a\, \mathcal{D}^{(1)}\right] .
\ee
Since we are considering an FLRW background metric, the shift $N^a=\barN^a+\delta N^a$ has zero $\barN^a$. This implies that the diffeomorphism constraint can be considered at the first order. 

Using the symmetry properties of
\eqref{Xijbca}
we can write the constraint for the gravitational part as
\be\kappa\;
\mathcal{D}_{g} =  \barp \, \partial_a \delta K^d_d 
- \barp \, \partial_d \delta K^d_a - \bark \, \partial_d \delta E^d_a~,
\ee
and for the matter part as
\be
\mathcal{D}_{m} = \barpi \, \partial_a \delta \varphi ~.
\ee
Recall that for tensor modes $\delta N^a =0$, therefore ${D}_{g}$ and ${D}_{m}$ play a role only for scalar and vector perturbations.

\paragraph*{Hamiltonian constraint--}
We consider the gravitational part of the Hamiltonian constraint up to the second order:
\be 
H[N] = \int_\Sigma d^3x \left[ \barn\,( \mathcal{H}^{(0)}+ \mathcal{H}^{(2)}) +  \delta N\, \mathcal{H}^{(1)}\right] .
\ee
 Using again the symmetry properties of
\eqref{Xijbca}, the expansion of the constraint given in \cite{lqc_review} gives 
\be
2 \kappa \; \mathcal{H}^{(0)} = - 6 \sqrt{\barp} \bark^2 ~, 
\ee
at  {\it zeroth order} 
and 
\be
2 \kappa\; \mathcal{H}^{(1)} = - 4 \sqrt{\barp} \delta K^d_d - \frac{\bark^2}{\sqrt{\barp}} \delta E^d_d + \frac{2}{\sqrt{\barp}}  \partial^j \partial_c \delta E^c_j,
\ee
at {\it first order},  for all kinds of perturbations. 
On the other hand, the {\it second order} turns out to be%
\ba\nonumber
2 \kappa\;  \mathcal{H}^{(2)} \!&\!=\!&
 - 2 \frac{\bark}{\sqrt{\barp}} \delta K^i_a\delta E^a_i \\\nonumber&& 
  \hspace{-1cm}
+ ~ \sqrt{\barp}\,  \big( \delta^b_i  \delta K^i_a \delta^a_j \delta K^j_b - \delta^a_i \delta K^i_a \delta^b_j \delta K^j_b \big)\,
 \lb{equationH21} \\\nonumber
&&   \hspace{-1cm}
  + ~ \frac{1}{4} \frac{\bark^2}{\barp^\frac{3}{2}}
  \big( \delta^i_a \delta E^a_i \delta ^j_b \delta E^b_j 
  - 2 \delta ^j_a \delta E^a_i \delta ^i_b \delta E^b_j 
  \big) \\\nonumber
&&   \hspace{-1cm}
 +~  \frac{1}{\barp^{\frac{3}{2}}} Y^{kjil}_{bdc} ~\epsilon^{ab}_k \; \partial_a \left( \delta E^d_j \partial_i \delta E^c_l\right) \lb{annulation} \\
&&  \hspace{-1cm}
 +~ \frac{1}{\barp^\frac{3}{2}} Z^{cidj}_{ab} \; \big( \partial_c \delta E^a_i\big)\big (\partial_d \delta E^b_j\big) \lb{BIGMAMMA}
\ea
and is different depending on the mode considered. 
The difference is only in the term $Z^{cidj}_{ab}$. Its explicit form reads 
\ba\lb{Zcidi}
Z^{cidj}_{ab} &=& \frac{1}{4} \epsilon^{ef}_k \epsilon^k_{mn} X^{mjd}_{be} X^{nic}_{af}  - \epsilon^{ie}_k X^{kjd}_{be} \delta^c_a 
\\\nonumber&&
- \epsilon^{ci}_k X^{kjd}_{ba} + \frac{1}{2}
\delta^i_a \epsilon^{ce}_k X^{kjd}_{be}. 
\ea
Imposing the conditions that define each mode and using \eqref{Xijbca}, we obtain that  the  term $Z^{cidj}_{ab} \; \big( \partial_c \delta E^a_i\big)\big (\partial_d \delta E^b_j\big)$ in \eqref{BIGMAMMA} is respectively
\ba
\hspace{-1cm}
&\delta_{ab} \delta^{ij} \delta^{cd}  \cdot ( \partial_c  \delta E^a_i )(\partial_d  \delta E^b_j) &  \text{for tensor modes,} \\
\hspace{-1cm}
&0&   \text{for vector modes,} \\
\hspace{-1cm}
&- \frac{1}{2} \delta^c_a \delta^d_b \delta^{ij} \cdot (\partial_c \delta E^a_i)( \partial_d \delta E^b_j)&  \text{ for scalar modes.}
\ea
This term is the only one that takes different forms when restricted to perturbations of the scalar, vector or tensor types. 
It follows that only the counterterms originating from this term will differ from one another for different types of perturbations.
\\

\paragraph*{\bf Quantum corrections--} In the classical case, the algebra is closed:
\\[-7mm]
\ba \lb{algtot} &\hspace{-9mm}
\{D_{(\!m\!+\!g\!)}[N^a_1],D_{(\!m\!+\!g\!)}[N^a_2]\} = 0 \,,
\hspace{-5mm} &\\ &\hspace{-9mm}
\{H_{(\!m\!+\!g\!)}[N],D_{(\!m\!+\!g\!)}[N^a]\} = - H_{(\!m\!+\!g\!)}[\delta N^a \partial_a \delta N] \,, 
\hspace{-5mm}& \\ &\hspace{-6mm}
\{H_{(\!m\!+\!g\!)}[N_1], H_{(\!m\!+\!g\!)}[N_2]\} = D_{(\!m\!+\!g\!)}\!\!\left[\!\frac{\barN}{\barp} \partial^a (\delta N_2 - \delta N_1) \!\right]\!.~
\lb{Dmg}
\ea
${D}_{g}$ does not undergo corrections from quantum effects \cite{Ashtekar:1995zh}.
We add quantum corrections at an effective level by replacing in the Hamiltonian constraint
\be
\bark \to  \frac{\sin ( \barmu \gamma \bark)}{ \barmu \gamma}
\ee
as a result of the quantization of the holonomies \cite{Ashtekar:2006wn}. The parameter $\barmu$, proportional to the ratio between the Planck length and the scale factor, carries the information on the scale at which quantum corrections become relevant.  This yields the quantum-corrected Friedmann equations
\be H^2=\frac\kappa3\rho\left(1-\frac\rho\rho_c\right)=\frac{\Hc^2}{\barp}\ee
where $H$ and $\Hc$ are the Hubble rate respectively in cosmic time and in conformal time, $\rho$ is the energy density and $\rho_c\approx0.4\,\rho_\Pl$ is the energy density at which a repulsive quantum-gravity force appears, removing the classical initial singularity \cite{lqc_review}.
The appearance of anomalies is contrasted by inserting counterterms in $\mathcal{H}^{(1)}$ and $\mathcal{H}^{(2)}$. For the explicit form of the resulting constraints, we refer the reader to the literature (see \cite{scalar}). The same modified constraints have been found in \cite{WilsonEwing:2011es}, where the counterterms of \cite{scalar} appears naturally after a Taylor expansion of the holonomies of the perturbed Ashtekar connection.

We are here interested in the structure of the resulting closed algebra. 
\\

\paragraph*{\bf Results--} Remarkably, the resulting quantum-corrected algebra valid for \emph{all}  different kind of perturbations is obtained with a single structure modification (\ref{algtot}-\ref{Dmg}). This appears in the last equations \eqref{Dmg}, which becomes
\ba
\{H_{(\!m\!+\!g\!)}[N_1], H_{(\!m\!+\!g\!)}[N_2]\} = {\bf \Omega} \; D_{(\!m\!+\!g\!)}\!\!\left[\frac{\barN}{\barp} \partial^a (\delta N_2 - \delta N_1) \right]
\nonumber\hspace{-5mm}&\\ \lb{Dmq}
\ea\\[-7mm]
where\\[-7mm] 
\begin{equation}
{\bf \Omega} = \cos(2\barmu \gamma \bark) = 1- 2 \frac{\rho}{\rho_c}~.
\label{omega}
\end{equation}
The single ${\bf \Omega}$ factor represents the quantum correction. It goes to 1 in the classical limit. This simple correction appears also in the definition of the evolution of the perturbations using gauge-invariant observables.

\paragraph*{Mukhanov-Sasaki equations of motion--} Finally, we can derive the correction to the Mukhanov-Sasaki \cite{Mukhanov:1990me} equation of motion for gauge-invariant perturbations of scalar and tensor type $v_{\mathrm{S(T)}}$. In conformal time, this is given by  \cite{scalar,Cailleteau:2011mi}
\begin{equation}\lb{EoM}
{v}''_{\mathrm{S(T)}} - {\bf \Omega} \, \nabla ^2 v_{\mathrm{S(T)}} - \frac{{z''_{\mathrm{S(T)}}}}{z_{\mathrm{S(T)}}} v_{\mathrm{S(T)}}= 0, 
\end{equation}
which reduces to the classical equation when ${\bf \Omega}\to1$. This equation holds for both scalar and tensor perturbations. Since we have considered the simple case of a scalar field, there is no vorticity and therefore there is no physical solution corresponding to vector perturbations.

For scalar perturbations, the Mukhanov variables in the quantum case are given   by
\be
v_{\mathrm{S}}=\sqrt{\barp}\,\left(\delta\phi + \frac{{\bar\varphi}'}{\Hc}\phi\right)
\hskip4mm \mbox{and}\hskip4mm
 z_{\mathrm{S}} \;=  
\; \sqrt{\barp}\, \frac{{\bar{\varphi}}'}{\Hc}.
\end{equation}

If we impose the divergence and the trace to be zero, we obtain 
for tensor modes 
\begin{equation}\lb{vzt}
v_{\mathrm{T}}=\; \sqrt{\frac{\barp}{{\bf \Omega}}} \; h
\hskip4mm \mbox{and}\hskip4mm
 z_{\mathrm{T}} \;=  
\; \sqrt{\frac{\barp}{{\bf \Omega}}}~,
\end{equation}
where $h$ represents the two degrees of freedom of $h^i_a$.
Inserting \eqref{vzt} into \eqref{EoM} we obtain the following form of the equations of motion for tensor perturbations:
\begin{equation} 
{h^i_a}'' + {h^i_a}'
 \; \left(2\Hc - {\frac{{\bf \Omega}'}{{\bf \Omega}}}\right) - {\bf \Omega} \, \nabla ^2 {h^i_a} = 0.
\end{equation}
This equation is clearly different from what has been used in previous works.
\\

\paragraph*{\bf Conclusions--}
We have presented a consistent framework for the study of perturbations in Loop Quantum Cosmology.
It is possible to write down a unique quantum-corrected algebra. This has a simple form, and the same quantum correction holds for all the different kinds of perturbations. 

This simple correction also appears in the Mukhanov-Sasaki equation, and consists in the insertion of the single factor \eqref{omega}, which becomes relevant only when the matter energy density approaches the Planck scale. 

We notice that there exist a small region in the strong quantum regime where ${\bf \Omega}$ becomes negative. This yields a
change of signature of the effective metric \cite{scalar,bojo_paily_2011} associated to the
appearance of divergences in the equation of motion of cosmological perturbations. As a consequence, new observable features could appear since the  value of tensor modes would be higher than in the classical case.
This have to be further investigated, possibly going beyond the effective treatment.

The existence of a single deformed closed algebra of constraints for all kind of perturbations, as exhibited in this work, is a strong case for the self-consistency of effective LQC.



\begin{thebibliography}{99}

\bibitem{lqg_review} 
C.~Rovelli, arXiv:1102.3660 [gr-qc];
C.~Rovelli, {\it  Quantum Gravity}, Cambridge, Cambridge University Press, 2004; 
T.~Thiemann, Lect. Notes Phys., 631, 41 (2003); T.~Thiemann, {\it Modern Canonical Quantum
General Relativity}, Cambridge, Cambridge University Press, 2007; 
R.~Gambini, J.Pullin, {\it A first course in Loop Quantum Gravity}, Oxford, Oxford University Press, 2011.

\bibitem{lqc_review} 
M.~Bojowald, Living Rev. Rel., 11, 4 (2008) [arXiv:gr-qc/0601085v1]; 
A.~Ashtekar \& P. Singh, Class. Quant. Grav. {\bf 28} (2011) 213001 [arXiv:1108.0893v2 [gr-qc]];
K.~Banerjee, G.~Calcagni, M.~Mart\'\i{n}-Benito, SIGMA 8 (2012), 016 [arXiv:1109.6801v2 [gr-qc]].

%
\bibitem{Langlois:1994ec}
  D.~Langlois,
  Class.\ Quant.\ Grav.\  {\bf 11} (1994) 389.

\bibitem{Bojowald:2008gz}
  M.~Bojowald, G.~M.~Hossain, M.~Kagan and S.~Shankaranarayanan,
  Phys.\ Rev.\  D {\bf 78} (2008) 063547
  [arXiv:0806.3929 [gr-qc]].
  
\bibitem{Bojowald:2008jv}
  M.~Bojowald, G.~M.~Hossain, M.~Kagan and S.~Shankaranarayanan,
  Phys.\ Rev.\  D {\bf 79} (2009) 043505
  [Erratum-ibid.\  D {\bf 82} (2010) 109903]
  [arXiv:0811.1572 [gr-qc]].
 
\bibitem{Bojowald:2007hv}
  M.~Bojowald and G.~M.~Hossain,
  Class.\ Quant.\ Grav.\, {\bf 24} (2007) 4801
  [arXiv:0709.0872 [gr-qc]].
  
\bibitem{Bojowald:2007cd}
  M.~Bojowald and G.~M.~Hossain,
  Phys.\ Rev.\  D {\bf 77} (2008) 023508
  [arXiv:0709.2365 [gr-qc]].

\bibitem{Bojowald:2010me}
  M.~Bojowald and G.~Calcagni,
 JCAP {\bf  1103} (2011) 32
 [arXiv:1011.2779 [gr-qc]].
 
\bibitem{Bojowald:2011hd}
  M.~Bojowald, G.~Calcagni and S.~Tsujikawa,
Phys. Rev. Lett. {\bf 107} (2011) 211302
  [arXiv:1101.5391 [astro-ph.CO]].

\bibitem{pheno_tensor}
J. Mielczarek, JCAP {\bf 0811} (2008) 011 [arXiv:0807.0712v1 [gr-qc]];
J. Grain and A. Barrau, Phys. Rev. Lett. {\bf 102} (2009) 081301;
J. Mielczarek, Phys. Rev. D {\bf 79} (2009) 123520 [arXiv:0902.2490v2 [gr-qc]]; 
J. Mielczarek, T. Cailleteau, J. Grain, A. Barrau, Phys. Rev. D {\bf 81} (2010) 104049 [arXiv:1003.4660v2 [gr-qc]];
J. Grain, A. Barrau, T. Cailleteau, J. Mielczarek, Phys. Rev. D {\bf 82} (2010) 123520 [arXiv:1011.1811v2 [astro-ph.CO]].
  
\bibitem{Li:2011zzd}
  Y.~Li and J.~Y.~Zhu,
  Class.\ Quant.\ Grav.\, {\bf 28} (2011) 045007
  [arXiv:1102.2720 [gr-qc]].

  %
\bibitem{Ashtekar:1995zh}
  A.~Ashtekar, J.~Lewandowski, D.~Marolf, J.~Mourao and T.~Thiemann,
  J.\ Math.\ Phys.\  {\bf 36} (1995) 6456
  [arXiv:gr-qc/9504018].
  


\bibitem{Ashtekar:2006wn}
A. Ashtekar, T. Pawlowski and P.Singh,
Phys. Rev. D {\bf 74} (2006) 084003 [arXiv:gr-qc/0607039v2].

\bibitem{vector}
J. Mielczarek, T. Cailleteau, A. Barrau, J. Grain, Class. Quant. Grav. {\bf 29} (2012) 085009 [arXiv:1106.3744v2 [gr-qc]].


\bibitem{scalar}
T. Cailleteau, J. Mielczarek, A. Barrau, J. Grain, Class. Quant. Grav. {\bf 29} (2012) 095010 [arXiv:1111.3535v2 [gr-qc]].

%
\bibitem{WilsonEwing:2011es}
  E.~Wilson-Ewing,
  Class.\ Quant.\ Grav.\  {\bf 29} (2012) 085005
  [arXiv:1108.6265 [gr-qc]].


\bibitem{Mukhanov:1990me}
  V.~F.~Mukhanov, H.~A.~Feldman and R.~H.~Brandenberger,
  Phys.\ Rept.\  {\bf 215} (1992) 203.
  
  %
\bibitem{Cailleteau:2011mi}
  T.~Cailleteau and A.~Barrau,
Phys. Rev. D {\bf 85} (2012) 123534
	[arXiv:1111.7192v1 [gr-qc]].

\bibitem{bojo_paily_2011}
	M.~Bojowald and G. M. Paily, arXiv:1112.1899v1 [gr-qc]
  
\end{thebibliography}
\end{document}